\documentclass[preprint,pra,superscriptaddress,showpacs,amsmath]{revtex4}

\usepackage{graphicx}

\begin{document}

\title
{Atom state evolution and collapse in ultracold gases during light scattering into a cavity}

\author{I. B. Mekhov}
\email{Mekhov@yahoo.com}
\affiliation{Department of Physics, Harvard University, Cambridge, Massachusetts, USA}

\author{H. Ritsch}
\affiliation{Institut f\"ur Theoretische Physik, Universit\"at
Innsbruck, Innsbruck, Austria}

\begin{abstract}
We consider the light scattering from ultracold atoms trapped in an optical lattice inside a cavity. In such a system, both the light and atomic motion should be treated in a fully quantum mechanical way. The unitary evolution of the light-matter quantum state is shown to demonstrate the non-trivial phase dependence, quadratic in the atom number. This is essentially due to the dynamical self-consistent nature of the light modes assumed in our model. The collapse of the quantum state during the photocounting process is analyzed as well. It corresponds to the measurement-induced atom number squeezing. We show that, at the final stage of the state collapse, the shrinking of the width of the atom number distribution behaves exponentially in time. This is much faster than the square root time dependence, obtained for the initial stage of the state collapse. The exponentially fast squeezing appears due to the discrete nature of the atom number distribution.
\end{abstract}


\maketitle

\newpage

\section{Introduction}

The interaction of quantized light with ultracold quantum gases represents a paradigmatic example of the ultimate quantum level of the light matter-interaction, where the quantizations of both light and matter (i.e. atomic motion) play equally important roles. In particular, a fully quantum description should be applied to describe both the state evolution and measurement process. The measurement of photons scattered from ultracold atoms provides an example, where the concept of quantum state collapse can be used to describe the measurement. Due to the recent experimental developments \cite{Exp1,Exp2,Exp3,Exp4,Exp5}, the realization of such schemes should become practical.

In our previous works, we considered the light scattering from atoms trapped in an optical lattice into a cavity. We have shown that the detection of photons leaking from the cavity represents a quantum nondemolition (QND) measurement of various atomic variables observing the light. Such a quantum measurement scheme can be used not only for the detection purposes \cite{NatPh,PRL07,PRA07,EPJD,LasPhys09}, but also for the active quantum state preparation, due to the light-matter entanglement and measurement back-action \cite{PRL09,PRA09,LasPhys10}.

In this paper, we present the details of the atomic state collapse during the photodetection process by analyzing the narrowing of the atom number distribution in time. We also comment on the short-time unitary evolution of the atomic state before the collapse process becomes important.

\section{Theoretical model}

We consider the model presented in Refs. \cite{PRL09,PRA09,LasPhys10}: $N$ ultracold atoms in an optical lattice of $M$ sites formed by strong off-resonant laser beams. A region of
$K\le M$ sites is illuminated by a weak external probe, which
is scattered into a cavity. Alternatively, this region is
illuminated by the cavity field appearing due to the presence of the
probe through the cavity mirror.

We use the open system approach for counting photons leaking the
cavity of decay rate $\kappa$. When a photon is detected, the jump
operator (the cavity photon annihilation operator $a_1$) is applied
to the quantum state: $|\Psi_c(t)\rangle \rightarrow
a_1|\Psi_c(t)\rangle$. Between the counts, the system evolves with a
non-Hermitian Hamiltonian. Such an evolution gives a quantum
trajectory for $|\Psi_c(t)\rangle$ conditioned on the detection of
photons.

The expression for the initial motional state of atoms reads
\begin{eqnarray}\label{1}
|\Psi(0)\rangle =\sum_{q}c_q^0 |q_1,..,q_M\rangle,
\end{eqnarray}
which is a superposition of Fock states reflecting all possible
classical configurations $q=\{q_1,..,q_M\}$ of $N$ atoms at $M$
sites, where $q_j$ is the atom number at the site $j$. As we have
shown in Refs. \cite{PRL09,PRA09}, the solution for conditional wave
function takes physically transparent form, if the following
approximations are used: atomic tunneling is much slower than light
dynamics and the probe waves are in the coherent state. The tunneling dynamics is indeed a very important process in many systems, e.g., in the double-well lattices \cite{Yukalov}. However, assuming here it is to be negligibly small, one can focus on a different type of the system dynamics: the state collapse, which exists even if the tunneling is neglected. The conditional state
after the time $t$ and $m$ photocounts at $t_1, t_2,..t_m$ is given by the quantum
superposition of solutions corresponding to the atomic Fock states
in Eq.~(\ref{1}):

\begin{eqnarray}\label{2x}
|\Psi_c(m,t)\rangle
=\frac{1}{F(t)}\sum_{q}\alpha_q(t_1)\alpha_q(t_2)...\alpha_q(t_m)
 \nonumber\\
\times  e^{\Phi_q(t)}c_q^0|q_1,...,q_M\rangle|\alpha_q(t)\rangle,
\end{eqnarray}
where
\begin{eqnarray}\label{3x}
\alpha_q(t)=\frac{\tilde{\eta}-iU_{10}\tilde{a}_0D^q_{10}}{i(U_{11}
D^q_{11}-\Delta_p)+\kappa}e^{-i\omega_pt}+ \nonumber \\
\left(\alpha_0 - \frac{\tilde{\eta}-iU_{10}
\tilde{a}_0D^q_{10}}{i(U_{11}
D^q_{11}-\Delta_p)+\kappa}\right)e^{-i(\omega_1+U_{11}D^q_{11})t-\kappa
t},
\end{eqnarray}
\begin{eqnarray}\label{4x}
\Phi_q(t)=\int_0^t\left[\frac{1}{2}(\eta\alpha^*_q-iU_{10}
a_0D^q_{10}\alpha^*_q-\text{c.c.})-\kappa|\alpha_q|^2\right]dt.
\end{eqnarray}

If the first photon is detected after the time $1/\kappa$, this solution simplifies further \cite{PRA09}:
\begin{eqnarray}
|\Psi_c(m,t)\rangle =\frac{1}{F(t)}\sum_{q}\alpha_q^m e^{\Phi_q(t)}
c_q^0 |q_1,...,q_M\rangle|\alpha_q\rangle, \label{2}\\
\alpha_q=\frac{\eta-iU_{10} a_0D^q_{10}}{i(U_{11}
D^q_{11}-\Delta_p)+\kappa}, \label{3}\\
\Phi_q(t)=-|\alpha_q|^2\kappa t+(\eta\alpha^*_q-iU_{10}
a_0D^q_{10}\alpha^*_q-\text{c.c.})t/2, \label{4}
\end{eqnarray}
where $\alpha_q$ is the cavity light amplitude corresponding to the
classical configuration $q$ (starting from here we use the slowly varying amplitudes). It is simply given by the Lorentz
function (\ref{3}) well-known from classical optics, where $a_0$ is
the external probe amplitude, $\eta$ is the amplitude of the probe
through a mirror; $U_{lm}=g_lg_m/\Delta_a$ ($l,m=0,1$), where
$g_{1,0}$ are the atom-light coupling constants, $\Delta_a=\omega_1
- \omega_a$ is the cavity-atom detuning; $\Delta_p=\omega_p -
\omega_1$ is the probe-cavity detuning. $D^q_{lm}=
\sum_{j=1}^K{u_l^*({\bf r}_j)u_m({\bf r}_j)q_j}$ are the
probe-cavity coupling coefficient and dispersive frequency shift
that sums contributions from all illuminated atoms with prefactors
given by the light mode functions $u_{0,1}({\bf r})$. Except the
prefactors associated with $m$ photodetections $\alpha_q^m$, the
components of quantum superposition in Eq.~(\ref{2}) acquires the
phases contained in $\Phi_q(t)$ (\ref{4}). $F(t)$ is the
normalization coefficient.

For several particular cases, the solution for the time-dependent
probability distribution of atoms corresponding to the state
(\ref{2}) can be simplified further \cite{PRL09,PRA09}. If the
probe, cavity, and lattice satisfy the condition of the diffraction
maximum for light scattering, the probability to find the atom
number $0<z<N$ in the lattice region of $K$ sites is given by
\begin{eqnarray}\label{5}
p(z,m,t)=z^{2m}e^{-z^2\tau}p_0(z)/\tilde{F}^2(m,\tau), \\
\tilde{F}^2(m,\tau)=\sum_z z^{2m}e^{-z^2\tau}p_0(z), \nonumber
\end{eqnarray}
with $\tau=2|C|^2\kappa t$, $C=iU_{10} a_0/(i\Delta_p-\kappa)$,
$p_0(z)$ is the initial distribution, and $\tilde{F}$ provides the
normalization. The light amplitude corresponding to the atom number
$z$ is $\alpha_z=Cz$.

If the condition of a diffraction minimum is satisfied, the
probability to find the atom number difference between the odd and
even sites in the lattice region of $K=M$ sites is given by the same
Eq.~(\ref{5}), but with a different meaning of the statistical
variable $-N<z<N$.

\section{State collapse for small atom number}

Equation~(\ref{5}) shows that any initially broad atom-number distribution $p_0(z)$ shrinks and approaches some very narrow distribution, which corresponds to the atom number squeezing. In Refs. \cite{PRL09,PRA09}, we have shown that the cental value of the final distribution $z_0$ is given by $z_{0}=\sqrt{m/\tau}$. Moreover, the full width at half maximum
(FWHM) of the distribution can be estimated as $\delta z \approx \sqrt{2\ln2/\tau}$, which shows the shrinking of the distribution in time \cite{PRL09,PRA09}. This type of shrinking (as $\sqrt{1/\tau}$) is rather typical for such photocounting schemes \cite{Onofrio}. This approximate formula works well if one assumes that (i) the distribution is already rather narrow, $\delta z \ll z_0$, but still (ii) the atom-number probabilities in Eq.~(\ref{5}) can be replaced by the continuous functions of $z$, which is a good approximation for very large atom numbers. For the large atom numbers, in Refs. \cite{LasPhys09} we have demonstrated that the state collapse can be very easily described analytically.

However, it is clear that while the distribution function continues to shrink, its width can reach the values of $\delta z \sim 1$, where the approximation of continuous functions obviously fails. In this case, one should explicitly take into account the discrete nature of the atomic ensemble. If one starts with a macroscopic quantum gas as in Refs. \cite{LasPhys09}, it is probably not very practical to expect that the distribution $p(z,m,t)$ can really shrink to the widths of $\delta z \sim 1$, because various distractive mechanisms will prevent the final state collapse to the many-body Fock state $|z_0, N-z_0\rangle$ with the precisely known atom number $z_0$ at the $K$ illuminated lattice sites and the rest $N-z_0$ atoms at the rest of $M-K$ sites. Instead, a distribution with some rather small $\delta z$ will establish. Thus the approximation $\delta z \approx \sqrt{2\ln2/\tau}$ can work well for the large atom numbers even till the last stage of the conditional time evolution. However, if the atom number and number of illuminated lattice sites are small, the situation $\delta z \sim 1$ is reasonable and practical experimentally \cite{GreinerMicroscope}.

We have carried out numerical simulations using the Quantum Monte Carlo Wave Function (QMCWF) simulations method \cite{PRL09,PRA09,LasPhys10}. The numerical results at a single quantum trajectory are presented in Fig. 1. They clearly show that the time evolution of the distribution function width has two different parts. First, the approximation of the square root decrease works very well. After that, however, the decrease and, thus the state collapse, becomes much faster. In the logarithmic scale figure, it is clear that the shrinking of the distribution becomes even exponential.

Such a behavior of the distribution function can be explained using the fact that the discrete functions of $z$ should be taken into account. Let us assume that the atom number $z$ changes discretely with a step $Z$. For example, for the atom number at $K$ sites, which can be measured at the directions of diffraction maxima, the step is $Z=1$ atom. In contrast, for the atom number difference between odd and even sites, which can be measured at the directions of diffraction minima, the step is $Z=2$ atoms, which reflects the total atom number conservation \cite{NatPh,PRL09,PRA09}. Note that, when the width of the distribution function is of the order of 1, introducing the FWHM becomes meaningless [e.g., the values of $p(z_0\pm Z)$ can be already less than $p(z_0)/2$]. Thus, the distribution width should be characterized directly by the square root of the atom number variance:

\begin{eqnarray}\label{6}
\langle z^2\rangle - \langle z\rangle^2 = \langle (z-\langle z\rangle)^2\rangle=\sum_z(z-\langle z\rangle)^2p(z).
\end{eqnarray}
The insight in the exponential shrinking can be made as follows. If one assumes that at the final stage of the state collapse only the atom numbers of $z_0$ and $z_0\pm Z$ are non-negligible, the Eq.~(\ref{6}) reduces to

\begin{eqnarray}\label{7}
\langle z^2\rangle - \langle z\rangle^2 \approx Z^2p(z_0-Z)+Z^2p(z_0+Z).
\end{eqnarray}
Using Eq.~(\ref{5}) (expanding the expressions in Taylor series in $Z/z_0$, taking into account that the normalization factor $\tilde{F}$ is a function of $m$, $\tau$, and $z_0$, and substituting $m$ as $m=z^2_0\tau$), one can get an estimation $\langle z^2\rangle - \langle z\rangle^2 \sim \exp{(-Z^2\tau)}$. This expression supports the exponential shrinking of the atom number distribution width at the final stage of the quantum state collapse, which is demonstrated in Fig. 1.

\section{Unitary dynamics of the quantum state}

In this section we address a question about the unitary (coherent) evolution of the light-matter state in the short-time limit, where the photon escapes are not important ($t\ll 1/\kappa$). We will be interested in the scattering into the direction of the diffraction maximum, where $D^q_{10}=N_K$ is the fluctuating atom number at $K$ lattice sites. We will consider only the transverse probe $a_0$ (while probing through the mirror does not present, $\eta=0$). The dispersion frequency shift $U_{11}D^q_{11}$ will be neglected in this configuration. Using the general solution (\ref{2x}-\ref{4x}) with those assamtions, and setting $\kappa=0$, $m=0$, one gets

\begin{eqnarray}\label{8}
|\Psi(t)\rangle
=\frac{1}{F(t)}\sum_{q}e^{\Phi_q(t)}c_q^0|q_1,...,q_M\rangle|\alpha_q(t)\rangle,
\end{eqnarray}
where
\begin{eqnarray}\label{9}
\alpha_q(t)= CD^q_{10}\left( 1-e^{i\Delta_pt}\right)e^{-i\omega_pt},
\end{eqnarray}
\begin{eqnarray}\label{10}
\Phi_q(t)=-i\Delta_p|CD^q_{10}|^2t+i|CD^q_{10}|^2 \sin\Delta_p t,
\end{eqnarray}
and $C=U_{10}a_0/\Delta_p$. For times such that $\Delta_pt \gg 1$, the phase gets even a simpler form: $\Phi_q(t)=-i\Delta_p|CD^q_{10}|^2t$. Thus, the evolution of the state for the interaction at the diffraction maximum reads:

\begin{eqnarray}\label{11}
|\Psi(t)\rangle
=\frac{1}{F(t)}\sum_{q}e^{-i\Delta_pC^2N_K^2t}c_q^0|q_1,...,q_M\rangle|\alpha_q(t)\rangle.
\end{eqnarray}

An important property of this solution is that the state components with various atom numbers at $K$ lattice sites have different phase evolutions, which quadratically depends on the atom number at $K$ sites $N_K$. For the superfluid state with the large atom number, if the number of illuminated sites is much smaller than the total site number, $K\ll M$, the statistics of $N_K$ is nearly Poissonian \cite{NatPh,PRL09,PRA09,LasPhys10}. Thus, the situation resembles the problems of a macroscopic BEC, where the collapse and revival of the matter field can be observed \cite{GreinerCollapse}. In both cases, the nontrivial dynamics is a consequence of the particular (i.e. quadratic in the atom number) modification of the phase evolution of the atom number components, initially constituting the Poissonian atom number distribution. The physical reasons of the phase modifications in two cases are however completely different. In the case of Ref. \cite{GreinerCollapse}, it is the atom-atom interaction that brings the nonlinearity in the problem in the form of the phase prefactor $\exp[{-iUN(N-1)t/2\hbar}]$, where $U$ is the atom-atom interaction energy. In our case, the atom-atom interaction has been completely neglected, and it is the atom-light interaction which leads to the phase dependence quadratic in $N_K$. As can be trace from the full solution (\ref{4x}), the phase depends on the product of probe-mode coupling coefficient proportional to $D^q_{10}$ (which is $N_K$ in this particular case), and the cavity mode amplitude $\alpha_q$. As we consider the light amplitudes as dynamical variables (and not as prescribed quantities as it is made in many problems of atoms in strong optical lattices), the light amplitudes are proportional to the atom number $N_K$. Thus, the product of the coupling coefficient and light amplitude gives us the term quadratic in the atom number $N_K$. In other words, one can say that the dynamical nature of the light mode leads to the "effective" atom-atom interaction.

It is probable, that similarly to Ref. \cite{GreinerCollapse}, one can obtain the collapse and revival of the matter field in our system, if some matter wave interference measurement will be carried out. Since we use the basis with the fixed total atom number $N$ (in contrast to Ref. \cite{GreinerCollapse}, where the coherent atomic state was assumed for the BEC), the calculation of the matter interference will be non-trivial. To calculate the interference pattern, one needs to include the atom counting procedure. This procedure is well-developed (see, e.g., Refs. \cite{Dalibard,Horak}) and leads to the results rather similar to what is expected from the coherent-state approximation for the initial atomic state. Besides that, the methods to disentangle light and matter can be used to observe the matter wave interference.

\section{Conclusions}

We considered the light scattering from ultracold atoms trapped in optical lattice inside a cavity. In such a system, both the light and atomic motion should be treated in a fully quantum mechanical way. The unitary evolution of the light-matter quantum state was shown to demonstrate the non-trivial phase dependence, quadratic in the atom number. This is essentially due to the dynamical self-consistent nature of the light modes assumed in our model. The collapse of the quantum state during the photocounting process has been analyzed as well. We have shown that, at the final stage of the state collapse, the shrinking of the width of the atom number distribution behaves exponentially in time. This is much faster than the square root time dependence, obtained for the initial stage of the state collapse. The exponentially fast atom number squeezing appears due to the discrete nature of the atom number distribution.

\section*{Acknowledgement}

The work was supported by the Austrian Science Fund FWF (project nos. J3005-N16 and S40130).

\newpage

Fig.1. Width of the atom number distribution function during the photodetection. Decreasing width corresponds to the atom number squeezing. At the initial stage, the shrinking is as $1/\sqrt{\tau}$, while at the final stage the shrinking is exponential. (a) Quantum trajectory without quantum jumps; (b) quantum trajectory with quantum jumps. Photodetection is at the diffraction minimum. Total number of atoms $N=100$, $K=M=100$ sites.

\begin{figure} \scalebox{1.7}[1.7]{\includegraphics{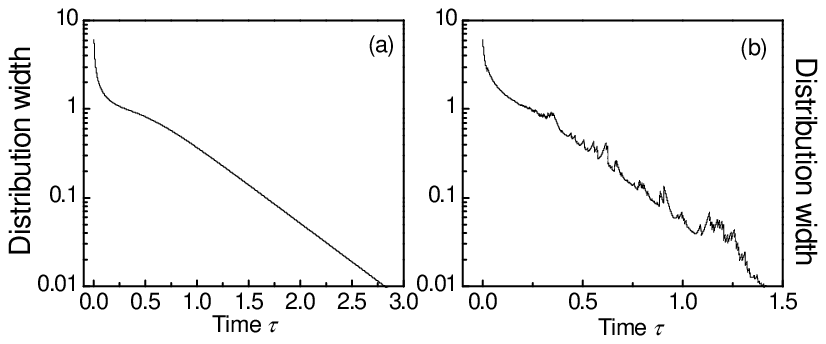}}
\caption{}
\end{figure}

\end{document}